\documentclass[twocolumn,tighten]{aastex61}
\usepackage{CJKutf8}
\usepackage{amsmath}
\usepackage{enumitem}
\usepackage{booktabs,pbox}
\usepackage{bm}

\graphicspath{{./figures/}{../figures/}}

\DeclareRobustCommand{\ion}[2]{%
\relax\ifmmode
\ifx\testbx\f@series
{\mathbf{#1\,\mathsc{#2}}}\else
{\mathrm{#1\,\mathsc{#2}}}\fi
\else\textup{#1\,{\mdseries\textsc{#2}}}%
\fi}

\def\hal [H$\alpha$]
\def\HI {\ion{H}{i}} 
\def\NII {\ion{N}{ii}} 

\newcommand{\Spitzer}{\it Spitzer}
\newcommand{\IRAS}{\it IRAS}
\def\2MASS {\it 2MASS}
\newcommand{\GALEX}{\it GALEX}
\def\m{$\mu$m}

\def\SDSS{{\it SDSS}}

\shorttitle{LVL SEDs}
\shortauthors{Dale/LVL}

\begin{document}


\title{Spectral Energy Distributions for 258 Local Volume Galaxies}



\newcommand{\UMass}   {\affiliation{Department of Astronomy, University of Massachusetts, Amherst, MA 01002, USA}}
\newcommand{\UWyoming}{\affiliation{Department of Physics and Astronomy, University of Wyoming, Laramie, WY 82071, USA}}
\newcommand{\UA}      {\affiliation{Centro de Astronomía (CITEVA), Universidad de Antofagasta, Avenida Angamos 601, Antofagasta, Chile}}
\newcommand{\NOIRLab} {\affiliation{Gemini Observatory/NSF’s NOIRLab, 950 N. Cherry Avenue, Tucson, AZ, 85719, USA}}
\newcommand{\UofA}    {\affiliation{Department of Astronomy and Steward Observatory, University of Arizona, Tucson, AZ, 85721, USA}}
\newcommand{\TAM}     {\affiliation{Department of Physics and Astronomy, Texas A\&M University, College Station, TX, 77843, USA}}
\newcommand{\AC}      {\affiliation{The Aerospace Corporation, El Segundo, CA, 90245 USA}}


\correspondingauthor{Daniel~A.~Dale}
\email{ddale@uwyo.edu}


\author[0000-0002-5782-9093]{Daniel~A.~Dale}
\UWyoming
\author[0000-0003-0946-6176]{M\'ed\'eric~Boquien}
\UA
\author[0000-0003-2261-5746]{Jordan~A.~Turner}
\UWyoming
\AC 
\author[0000-0002-5189-8004]{Daniela~Calzetti}
\UMass
\author[0000-0001-5448-1821]{Robert~C.~Kennicutt, Jr.}
\UofA
\TAM
\author[0000-0003-0946-6176]{Janice~C.~Lee}
\NOIRLab


\begin{abstract}
We present model spectral energy distribution (SED) fits to ultraviolet/optical/infrared observations for the 258 nearby galaxies in the Local Volume Legacy survey, a sample dominated by lower-luminosity dwarf irregular systems.  The data for each galaxy include up to 26 spatially-integrated broadband and narrowband fluxes from the Galaxy Evolution Explorer, Spitzer Space Telescope, and Infrared Astronomical Satellite space-based platforms and from the Sloan Digital Sky Survey, Two Micron All Sky Survey, and other ground-based efforts.  The CIGALE SED fitting package is employed using a delayed star formation history with an optional late burst or quenching episode to constrain 11 different free parameters that characterize the properties of each galaxy's stellar and dust emission, with the overriding constraint that the ultraviolet/optical emission absorbed by interstellar dust grains is emitted in equal energy portions at infrared wavelengths.  The main results are: i) 94\% of the SED fits yield reduced $\chi^2$ values less than 3; ii) the modeled stellar masses agree with those derived from 3.6~\m-based measures with a scatter of 0.07~dex; iii) for a typical galaxy in the sample the SED-derived star formation rate averaged over the past 100~Myr is about 88\% of the value derived from standard hybrid indicators on similar timescales; and iv) there is a statistically significant inverse relation between the stellar mass fraction appearing in the late burst and the total stellar mass.  These results build upon prior SED modeling efforts in the local volume and lay the groundwork for future studies of more distant low-metallicity galaxies with JWST.

\end{abstract}




\section{Introduction} \label{sec:intro}
Understanding the interplay between star formation and the interstellar medium within galaxies has long been a primary goal of galaxy evolution studies.  However, while the broad brush strokes of star formation have come into focus over the past decades, many details remain to be filled in.  Infrared studies of nearby galaxies provide a two-fold advantage to understanding the processes that underlie the evolution of star formation and the interstellar medium: enhanced spatial resolution and the ability to penetrate much of the gas and dust that otherwise obscure ultraviolet and optical observations.  Several studies of relatively large samples of nearby galaxies have utilized space-based infrared observatories, e.g., the Spitzer Nearby Galaxies Survey \citep[SINGS;][]{kennicutt2003}, the Great Observatories All-Sky LIRG Survey \citep[GOALS;][]{armus2009}, the Spitzer Survey of Stellar Structure in Galaxies \citep[S$^4$G;][]{sheth2010}, Key Insights on Nearby Galaxies: A Far-Infrared Survey with Herschel \citep[KINGFISH;][]{kennicutt2011}, the Dwarf Galaxy Survey \citep[DGS;][]{madden2013}, and A Survey of Far-Infrared Lines in Nearby Galaxies \citep[SHINING;][]{herrera-camus2018}.  But none of these surveys provides a statistically representative survey of the population of galaxies found in the local volume.  Most of these efforts are biased toward massive, metal-rich, and high surface brightness systems. The sampling that does exist for lower mass systems is sparse, and is far from representative, despite the fact that this population offers the greatest diversity of properties, the best-measured star formation histories, and hence optimal leverage for elucidating the processes that underlie star formation and shape the properties of galaxies.

The Local Volume Legacy (LVL) survey has provided an unbiased, fully representative, and statistically robust sample of nearby star-forming galaxies.  The overarching goal of the LVL survey has been to provide critical insight into two of the primary processes that shape the growth of galaxies: star formation and its interaction with the interstellar medium.  In a series of papers the LVL team has investigated the spatially-resolved star formation, dust, and red stellar populations of local galaxies which span the full diversity of luminosities, surface brightnesses, metallicities, dust characteristics, and star formation properties \citep[e.g.,][]{lee2009,boquien2010,calzetti2010,marble2010,bothwell2011,berg2012,weisz2012,grasha2013,cook2014a,cook2016}.  The tiered survey includes: (1) all known galaxies inside a sub-volume bounded by 3.5~Mpc, and (2) an unbiased sample of galaxies within the larger, and more representative, 11~Mpc sphere.  This strategy provides volume-complete coverage of galaxies over the entire luminosity function, with the minimum sample needed to fully characterize the local galaxy population.  The data backbone of the LVL survey has been the Galaxy Evolution Explorer ({\it GALEX}), H$\alpha$ narrowband, and Spitzer Space Telescope imaging obtained for the volume-complete sample.  Additional broadband optical and infrared observations have been obtained or compiled to complete the panchromatic census.

In this paper we utilize the CIGALE software package to fit the multi-wavelength dataset to constrain the physical properties and star formation histories of the LVL sample.  \cite{janowiecki2017} previously explored spectral energy distribution (SED) fitting of the LVL sample in the context of using it as a reference sample for their study of 18 blue compact dwarf galaxies.  \cite{janowiecki2017} used CIGALE to fit the LVL SEDs, utilizing 14 different broadband fluxes from {\it GALEX} FUV/NUV, $UBVR_{\rm C}I_{\rm C}$, 2MASS $JHK_{\rm s}$, {\it Spitzer} IRAC, and {\it Spitzer} MIPS 24~\m.  Their main finding was no surprise: compared to the overall dwarf galaxy population, blue compact dwarf galaxies are currently undergoing extreme star formation episodes, as quantified by the relative masses and amplitudes of the starbursts.  We expand beyond the work of \cite{janowiecki2017} by including up to 26 fluxes for the LVL sample---the fluxes employed in \cite{janowiecki2017} plus H$\alpha$ narrowband, SDSS $ugriz$, {\it Spitzer} MIPS 70 and 160~\m, and {\it IRAS} broadband fluxes, where available.  In addition, our fits utilize the \cite{draine2014} dust emission models instead of the one-parameter dust models of \cite{dale2014}, so we are able to report physical properties of the dust population such as the dust mass fraction that is locked up in polycyclic aromatic hydrocarbons (PAHs).  We also use a ``delayed'' star formation model with the possibility of a late quenching or burst episode, whereas they used an exponentially decaying SFH with an exponentially-decaying late burst.  Finally, \cite{janowiecki2017} only reported the fit parameters for their sample of 18 blue compact dwarf galaxies; we report here the fit parameters for the larger LVL sample.

Section~\ref{sec:sample} summarizes the sample of galaxies studied in this effort.
Section~\ref{sec:data} reviews and presents the multi-wavelength data set used here, and the analysis of the data is presented in Section~\ref{sec:analysis}.
Section~\ref{sec:results} presents the main results, and Section~\ref{sec:summary} provides concluding remarks.


\section{Sample} \label{sec:sample}
\begin{figure}
 \plotone{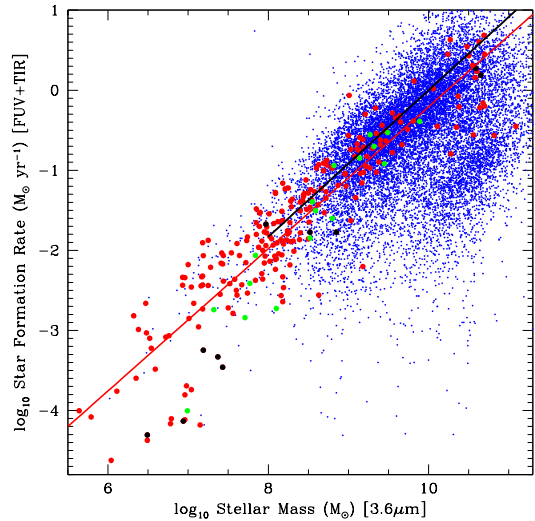}
 \caption{Comparison of global star formation rates and stellar masses for the LVL sample (filled red/black/green circles) with the ``z0MGS'' $z=0$ Multiwavelength Galaxy Synthesis catalog (blue dots) of \cite{leroy2019}.  The star-forming main sequence linear relation from \cite{peng2010} is shown as a black line; our best-fit relation for the LVL sample is provided as a red line.  LVL star formation rates are based on the \cite{hao2011} FUV+TIR calibrations for 216 galaxies (red circles), the \cite{hao2011} NUV+TIR calibrations for 11 galaxies without (GALEX) FUV observations (magenta circles), and the \cite{kennicutt2009} H$\alpha$+24~\m\ calibration for 17 galaxies with neither (GALEX) NUV nor FUV observations (green circles).  Stellar masses are based on 3.6~\m\ fluxes using the prescription outlined in Appendix~A.4 of \cite{leroy2019}.}
 \label{fig:greenvalley}
\end{figure}

The Local Volume Legacy sample of 258 galaxies is fully described in \cite{dale2009}.  Briefly, the sample has a two-fold lineage that derives from the ``ANGST'' ACS Nearby Galaxy Survey Treasury \citep{dalcanton2009} and the ``11HUGS'' 11~Mpc H$\alpha$ and Ultraviolet Survey \citep{kennicutt2008,lee2009}.  ANGST is a volume-limited survey beyond the Local group that reaches out to 3.5~Mpc and additionally samples the M~81 group and Sculptor filament.  The 11HUGS sample targeted star-forming disk galaxies out to $D\leq 11$~Mpc that met the criterion $m_B<15$~mag.  Both of these surveys avoid the Galactic plane, with the criterion for ANGST satisfying $|b|>20^{\circ}$ and that for 11HUGS following $|b|>30^{\circ}$.  Figure~\ref{fig:greenvalley} demonstrates the sample's range of star formation rates and stellar masses based on 3.6~\m\ luminosities.  Clearly this volume-limited survey provides a different sampling of the galaxy population than traditional notions of the ``red sequence'' and ``blue cloud''; a large fraction of the LVL sources skews to much smaller stellar masses and star formation rates than the bulk of the 15,750 objects in the ``z0MGS'' $z=0$ Multiwavelength Galaxy Synthesis catalog of nearby ($\lesssim50$~Mpc) galaxies \citep{leroy2019}.  In fact, the z0MGS sample targets galaxies more luminous than the Large Magellanic Cloud (LMC) by adopting a luminosity cut of $M_B<-18$~mag and 81\% of the LVL sample is less luminous than the LMC \citep{dale2009}.  If we define a dwarf galaxy as having a stellar mass less than $10^9~M_\odot$, then 75\% of the LVL sample is populated by dwarf systems.  This deviation of the LVL sample from the red-and-dead and main sequence norms is reflected in the distribution of optical morphologies: the sample is comprised of 8\%, 31\%, and 60\% early-type/lenticular, spiral, and irregular galaxies, respectively.  

The median distance for the sample is 5.9~Mpc.  The majority of the distances lie between 0.5~Mpc and 11~Mpc, with outliers being the Magellanic Clouds (50--60~kpc), UGC~06782 (14~Mpc), IC~2049 (17~Mpc), and UGC~07321 (20~Mpc).  Most of the distances derive from either tip of the red giant branch measurements (39.9\%) or from flow-corrected velocities (35.3\%).  The distance estimates for the remaining galaxies derive from an assortment of techniques including the luminosity of the brightest stars (13.2\%), Cepheid observations (6.2\%), surface brightness fluctuations (3.1\%), membership in a known galaxy group (1.6\%), and the Tully-Fisher relation (0.8\%).  Additional details on the galaxy distances may be found in \cite{dale2009}.

\section{Data} \label{sec:data}
The Local Volume Legacy broadband datasets used in this work include {\it GALEX} ultraviolet \citep{lee2011}, 2MASS near-infrared \citep{dale2009}, and {\it Spitzer} infrared imaging \citep{dale2009}.  We also utilize the spatially-integrated optical broadband flux densities provided by \cite{cook2014b}, who use the same photometric apertures as in \cite{dale2009} (see also \citealt{dale2007,dale2012,dale2017}).  As explained in \cite{dale2009}: i) the photometric apertures have been carefully crafted to capture all of the observable emission, ii) the apertures are the same across all optical/infrared wavelengths, and iii) the appropriate aperture corrections have been applied at each wavelength.  Additional standard corrections to the flux densities used in the SED fitting, e.g., removing the local ``sky'' / foreground stars / background galaxies and correcting for foreground Milky Way extinction, have been applied as well\footnote{However, we point out that the fluxes provided in the Appendix are not corrected for Milky Way foreground extinction.}.  We have supplemented these datasets with global {\it IRAS} infrared flux densities taken from \cite{rice1988}, \cite{moshir1990}, \cite{sanders2003}, and our own extractions using the IRAS Scan Processing and Integration Tool\footnote{https://irsa.ipac.caltech.edu/IRASdocs/scanpi\_over.html}.  The majority of the LVL galaxies have infrared surface brightnesses that are too faint for secure {\it IRAS} detections; including our own SCANPI-based extractions for [6,~8,~23,~23] LVL galaxies with no published {\it IRAS} measurements, we have global {\it IRAS} flux densities for [61,~70,~125,~119] galaxies, respectively, for the IRAS [12~\m,~25~\m,~60~\m,~100~\m] filters.  A comparison of the 8 new SCANPI-based {\it IRAS} 25~\m\ fluxes with the corresponding {\it Spitzer} 24~\m\ fluxes shows a median flux ratio of $f_\nu$(25~\m)/$f_\nu$(24~\m)=1.2 and a dispersion of 0.1~dex.  The major axes of these 23 galaxies with new SCANPI-based {\it IRAS} fluxes are all larger than the angular resolution of {\it IRAS}.  Finally, we include integrated H$\alpha$ fluxes that are corrected for \NII\ emission from \cite{kennicutt2008}.  The table of photometry is provided in the Appendix.  Table~\ref{tab:detections} provides the number of available detections as well as the number of upper limits for each of the 26 wavelengths employed here for the overall sample of 258 LVL galaxies.  The treatment of upper limits in the SED fitting is described in Section~4.3 of \cite{boquien2019}, which follows from the technique laid out in Appendix~A2 of \cite{sawicki2012}.  The upper limits and less secure photometry are limited to low-mass galaxies; 91\% (100\%) of the upper limits correspond to galaxies with $\log_{10} M_*/M_\odot \lesssim 8.0$ (8.7).

\begin{deluxetable}{llrrr}
\tabletypesize{\scriptsize}
\tablecaption{Detections and Upper Limits by Wavelength\label{tab:detections}}
\tablewidth{0pc}
\tablehead{
\colhead{Telescope} &
\colhead{Filter} &
\colhead{$\lambda$} &
\colhead{\# detections} &
\colhead{\# upper}
\\
\colhead{} &
\colhead{} &
\colhead{(\m)} &
\colhead{} &
\colhead{~~~limits}
}
\startdata
\hline
\hline
\GALEX  & $FUV$      & 0.15 & 222 & 0 \\ 
\GALEX  & $NUV$      & 0.23 & 233 & 0 \\ 
multiple& $U$        & 0.36 & 132 & 0 \\ 
multiple& $B$        & 0.44 & 185 & 0 \\ 
multiple& $V$        & 0.55 & 129 & 0 \\ 
multiple& $R_{\rm C}$& 0.71 & 188 & 0 \\ 
SDSS    & $u$        & 0.29 & 144 &  1 \\ 
SDSS    & $g$        & 0.48 & 146 & 0 \\ 
SDSS    & $r$        & 0.62 & 146 & 0 \\ 
SDSS    & $i$        & 0.76 & 146 & 0 \\ 
SDSS    & $z$        & 0.91 & 146 & 0 \\ 
2MASS   & $J$        & 1.24 & 229 & 26 \\ 
2MASS   & $H$        & 1.66 & 228 & 27 \\ 
2MASS   & $K_{\rm s}$& 2.16 & 219 & 36 \\ 
\Spitzer& IRAC1      & 3.6  & 247 &  9 \\ 
\Spitzer& IRAC2      & 4.5  & 247 & 10 \\ 
\Spitzer& IRAC3      & 5.8  & 212 & 43 \\ 
\Spitzer& IRAC4      & 8.0  & 214 & 43 \\ 
\Spitzer& MIPS1      &  24  & 210 & 47 \\ 
\Spitzer& MIPS2      &  70  & 204 & 53 \\ 
\Spitzer& MIPS3      & 160  & 198 & 59 \\ 
\IRAS   & IRAS1      &  12  &  61 & 0 \\ 
\IRAS   & IRAS2      &  25  &  70 & 0 \\ 
\IRAS   & IRAS3      &  60  & 125 & 0 \\ 
\IRAS   & IRAS4      & 100  & 119 & 0 \\ 
multiple&H$\alpha$   & 0.6563 & 218 & 6 \\
\hline
\enddata
\end{deluxetable}


\section{Spectral Energy Distribution Fits} \label{sec:analysis}
We employ the CIGALE spectral energy distribution fitter \citep{boquien2019} to constrain several physical parameters that characterize the stellar and dust properties in the LVL galaxies.  CIGALE functions on a principle of energy balance, whereby the (primarily) ultraviolet and optical stellar emission that is absorbed by interstellar dust grains reappears in the infrared in energetically equal amounts in the form of dust continuum(s) and PAH feature emission \cite[see also, e.g.,][]{dacunha2008,leja2017}.  Both nebular and synchrotron emission are also modeled in CIGALE.  Our adoptions include the stellar and dust emission libraries of \cite{bruzual2003} and \cite{draine2014}, the \cite{chabrier2003} stellar initial mass function, and a dust attenuation curve based on \cite{calzetti2000} and \cite{leitherer2002}.  The grid values for each parameter in the fits are listed in Table~\ref{tab:parameters}.  The total number of grid points (or different models) is $4 \times 7 \times 5 \times 14 \times 18 \times 3 \times 18 \times 6 \times 9 \times 3 \times 11 = 3.4\cdot 10^9$.

We follow \cite{hunt2019} and opt for a ``delayed'' star formation history with the option for an additional late burst or quenching of the star formation that begins at time $t_{\rm bq}$, also known as the \texttt{sfhdelayedbq} module within CIGALE:
\begin{equation}
{\rm SFR}(t) \propto\ \begin{cases}
t \exp(-t/\tau_{\rm main}), & \text{$t\leq t_{\rm bq}$} \\
r_{\rm SFR} \  {\rm SFR}(t\,=\,t_{\rm bq}), & \text{$t> t_{\rm bq}$} \quad .
\end{cases}
\label{eqn:sfr}
\end{equation}
Previous CIGALE-based efforts for samples of nearby star-forming galaxies have shown that there are minimal differences ($\sim0.1$~dex) in the output fit values when using a delayed exponential $+$ late burst star formation history versus using a purely exponentially-decaying $+$ late burst \citep{dale2016,salim2016}.  However, we find here for the LVL sample that implementing a late burst/quench using the \texttt{sfhdelayedbq} CIGALE module provides superior SED fits (i.e., lower $\chi^2_{\rm reduced}$) than when invoking the \texttt{sfhdelayed} module with a late burst option.  The latter module yields poor fits in the ultraviolet wavelength range for about a third of the LVL sample, which results in SED-based star formation rates and stellar masses that are in poor agreement with those derived by independent measures that are not based on SED fits (see \S~\ref{sec:analysis}).  Two of the parameters relevant to our adopted star formation history are the age and $e$-folding time $\tau_{\rm main}$ of the main stellar population.  The onset age for the main stellar population is fixed to 13~Gyr but the results of this work are insensitive to values tested between 10 and 13~Gyr.  We also model a possible late quenching or burst event via its age ($t_{\rm bq}$ in Equation~\ref{eqn:sfr} is when the burst/quench turns on; \texttt{age\_bq} in Table~\ref{tab:parameters} is the duration of the burst/quenching) and a multiplicative amplitude $r_{\rm SFR}$ that applies instantaneously at the time of the burst / quenching event (see \citealt{boquien2019} for details).   Though it is tempting to tailor the star formation history separately for different galaxy bins (e.g., optical morphology or stellar mass), for ease of comparison we adopt the same star formation history for all galaxies and assume identical grid values for each SED fit.  

The modeled attenuation curve allows for ranges in the color excess for both nebular line and stellar continuum emission and the slope $\delta$ of a power law that modifies the curve.  The UV ``bump'' amplitude is fixed to be zero.  The properties of the dust emission that are allowed to vary include the PAH mass fraction $q_{\rm PAH}$, the minimum intensity $U_{\rm min}$ of the radiation field that illuminates the dust, and the fraction of the dust mass $\gamma$ that is associated with the power-law (i.e., non-diffuse) part of the starlight intensity distribution \citep[see][]{draine2014}.
Nebular emission is included in the spectral energy distribution fits, with the chosen (fixed) parameter values as listed in Table~\ref{tab:parameters}.  The output parameter values and uncertainties from the spectral energy distribution fits come in two flavors: a ``best-fit'' solution via a $\chi^2$ minimization as well as a Bayesian-like approach that provides output based on a likelihood-weighting scheme involving the $\chi^2$ value for each model grid point \cite[\S~4.4 of][]{boquien2019}.  In almost all cases we utilize the fit values derived from the more robust Bayesian-based process, in order to avoid the limitations imposed by the discreteness of the grid-based best-fit solutions \citep{taylor2011}.

\begin{deluxetable*}{lll}
\tabletypesize{\scriptsize}
\tablecaption{CIGALE SED Fit Parameters\label{tab:parameters}}
\tablewidth{0pc}
\tablehead{
\colhead{Parameter} &
\colhead{Notation} &
\colhead{Allowed Values} 
}
\startdata
\hline
\multicolumn{3}{c}{\texttt{bc03} Bruzual \& Charlot (2003) stellar library} \\
\hline
Metallicity	                                 &\texttt{metallicity}& 0.004, 0.008, 0.02, 0.05\\
Initial Mass Function	                     &\texttt{imf}        & 1 (Chabrier 2003)\\ 
\hline
\multicolumn{3}{c}{\texttt{sfhdelayedbq} delayed star formation history with optional constant burst/quench} \\
\hline
$e$-folding time of the main population (Gyr)&\texttt{tau\_main}  & 1, 2, 3, 4, 5, 7.5, 10\\
Age of oldest stars (Gyr)                    &\texttt{age\_main}  & 13\\	
Age of the burst/quench episode (Myr)        &\texttt{age\_bq}    & 10, 25, 50, 75, 100\\	
Ratio of the SFR after/before \texttt{age\_bq} &\texttt{r\_sfr}   & 0, 0.2, 0.4, 0.6, 0.8, 1, 1.25, 1.5, 1.75, 2, 2.5, 5, 7.5, 10\\	
\hline
\multicolumn{3}{c}{\texttt{nebular} nebular emission} \\
\hline
Ionization parameter                         &\texttt{logU}       & $-$3.0\\
Electron density (cm$^{-3}$)                 &\texttt{n\_e}       &100\\
Lyman continuum escape fraction              &\texttt{f\_esc}     & 0.0\\
Lyman continuum dust absorption fraction     &\texttt{f\_dust}    & 0.0\\
Emission line width (km~s$^{-1}$)            &\texttt{lines\_width}& 300\\
\hline
\multicolumn{3}{c}{\texttt{dustatt\_modified\_starburst} attenuation curve} \\
\hline
Color excess for nebular emission            &\texttt{E\_BV\_lines}& 0.01,0.02,0.04,0.06,0.08,0.1,0.15,0.2,0.25,0.3,0.35,0.4,0.5,0.6,0.7,0.8,0.9,1\\
Reduction factor applied to \texttt{E\_BV\_lines} for stellar $E(B-V)$ & \texttt{E\_BV\_factor}& 0.25, 0.50, 0.75	 \\
UV bump amplitude                       &\texttt{uv\_bump\_amplitude}& 0\\
Slope $\delta$ of power law that modifies attenuation curve&\texttt{powerlaw\_slope}&[$-1.5$,0.2] with spacings of 0.1\\
\hline
\multicolumn{3}{c}{\texttt{dl2014} dust emission} \\
\hline
PAH dust mass fraction $q_{\rm PAH}$    &\texttt{qpah}& 0.47, 1.12, 2.50, 3.90, 5.26, 6.63\\
Minimum radiation field $U_{\rm min}$   &\texttt{umin}& 0.1, 0.25, 0.50, 1.0, 2.5, 5.0, 10.0, 25.0, 50.0\\
Power law slope $dM_{\rm dust}/dU \propto U^{-\alpha}$&\texttt{alpha}& 1.5, 2.0, 2.5\\
Illumination fraction from $U_{\rm min}$ to $U_{\rm max}$&\texttt{gamma}& [$10^{-3}$,$10^{-0.5}$] with spacings of 0.25~dex\\ 
\hline
\enddata
\end{deluxetable*}

The resulting SED fit parameter values are presented in the Appendix for each galaxy.  For galaxies without secure MIPS~70\m\ and 160\m\ detections, we provide dust mass upper limits as the sum of the SED-derived dust mass plus its Bayesian uncertainty.  Accordingly, for these galaxies the interpretation of the other dust-related parameters is limited.


\section{Results} \label{sec:results}
\begin{figure}
 \plotone{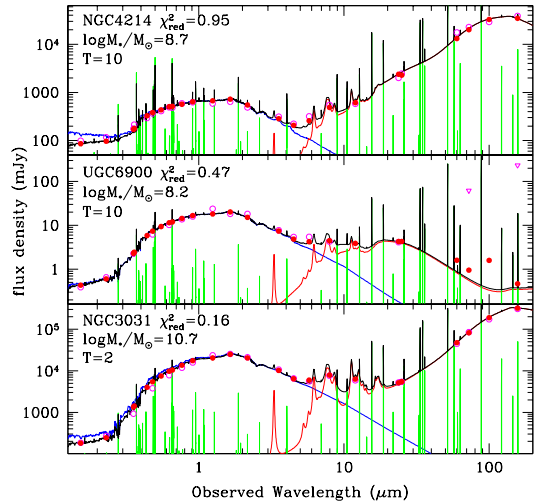}
 \caption{Example CIGALE fits for three LVL galaxies, one a massive spiral galaxy (NGC~3031 = M~81) and two lower luminosity irregular galaxies (NGC~4214 and UGC~6900).  The blue, red, green, and black lines represent the intrinsic stellar continuum, dust continuum, nebular, and total modeled emission, respectively.  The filled red circles represent the model flux densities in the 25 possible broadband bandpasses whereas the open magenta circles are the observed flux densities (upper limits are open magenta triangles).}
 \label{fig:sed}
\end{figure}

Figure~\ref{fig:sed} provides example CIGALE spectral energy distribution fits for two low-luminosity irregulars and one massive spiral galaxy in the LVL sample.  The differences between the observed and intrinsic ultraviolet/optical spectra are due to absorption by dust and are equal in energy to that emitted by dust in the infrared, as prescribed by design in the fitting code.  The faintest target of these three example galaxies, UGC~6900, lacks infrared detections at wavelengths longward of 24~\m\ (see Table~\ref{tab:detections}) and thus the SED fit for this lower-luminosity galaxy relies on a more limited wavelength range than do the fits for NGC~4214 and NGC~3031 (we have incorporated the flux upper limits in our CIGALE SED fitting).  The reduced $\chi^2$ values for these three example cases are all less than unity.  

\begin{figure}
 \plotone{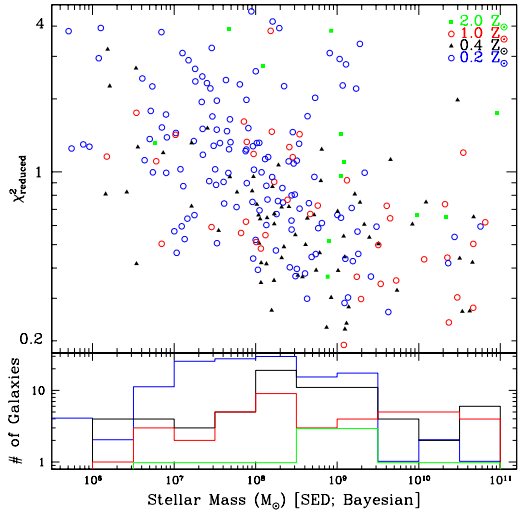}
 \caption{Top: The reduced $\chi^2$ values as a function of SED-derived stellar mass.  The points are color-coded according to best-fit stellar metallicity.  Bottom: The distributions of the best-fit metallicities as a function of stellar mass using the same color scheme as in the top panel.}
 \label{fig:p2boquien}
\end{figure}

The top panel in Figure~\ref{fig:p2boquien} displays $\chi^2_{\rm reduced}$ for the entire sample as a function of total SED-derived stellar mass.  There is a trend with lower-mass (fainter) galaxies having larger $\chi^2_{\rm reduced}$, as expected (the Spearman's rank correlation coefficient is $-$0.51).  The histograms in the bottom panel of Figure~\ref{fig:p2boquien} support a scenario of higher best-fit metallicities for more massive galaxies, also in agreement with expectations \citep[e.g.,][]{tremonti2004}.  The median $\chi^2_{\rm reduced}$ for the CIGALE fits to the LVL sample is 0.81 with a $1\sigma$ dispersion of 0.41, reflecting that the SED fits are generally reasonable for the sample.  Moreover, given that most $\chi^2_{\rm reduced}$ values are less than unity, the distribution suggests that the typical photometric uncertainties are overestimated.  However, our $\chi^2_{\rm reduced}$ are driven to artificially lower values since we incorporate a 10\% across-the-board increase in our flux uncertainties (\texttt{additionalerror}$=$0.1) in our SED fitting.  If we instead reduce the additional flux uncertainty to 5\%, the median $\chi^2_{\rm reduced}$ increases to 1.54 and the scatter grows to 0.77.  Ninety-four percent of the SED fits yield $\chi^2_{\rm reduced}<3$ and all but four fits ($>98$\%) result in $\chi^2_{\rm reduced}<4$.  The galaxies with $\chi^2_{\rm reduced}>3$ are generally low mass galaxies ($\log_{10} M_*/M_\odot \lesssim 8$) with suspect photometry.  Though there is a large fraction of galaxies (51\%) with a best-fit stellar metallicity of $Z=0.004$ (1/5$^{\rm th}Z_\odot$ like the SMC), these are nearly all dwarf galaxy systems with $M_* < 10^9~M_\odot$ and thus are expected to have significantly sub-solar metal abundances \citep{lee2006,kirby2013}.  The remaining systems have best-fit metallicities of 0.4$Z_\odot$ (27\%), $Z_\odot$ (16\%), and $Z=2.5Z_\odot$ (5\%).  We note that there is no trend in $\chi^2_{\rm reduced}>3$ with the number of SED data points used in the fit.

\begin{figure*}
 \includegraphics[height=13cm,trim={0 4.2cm 0 0},clip]{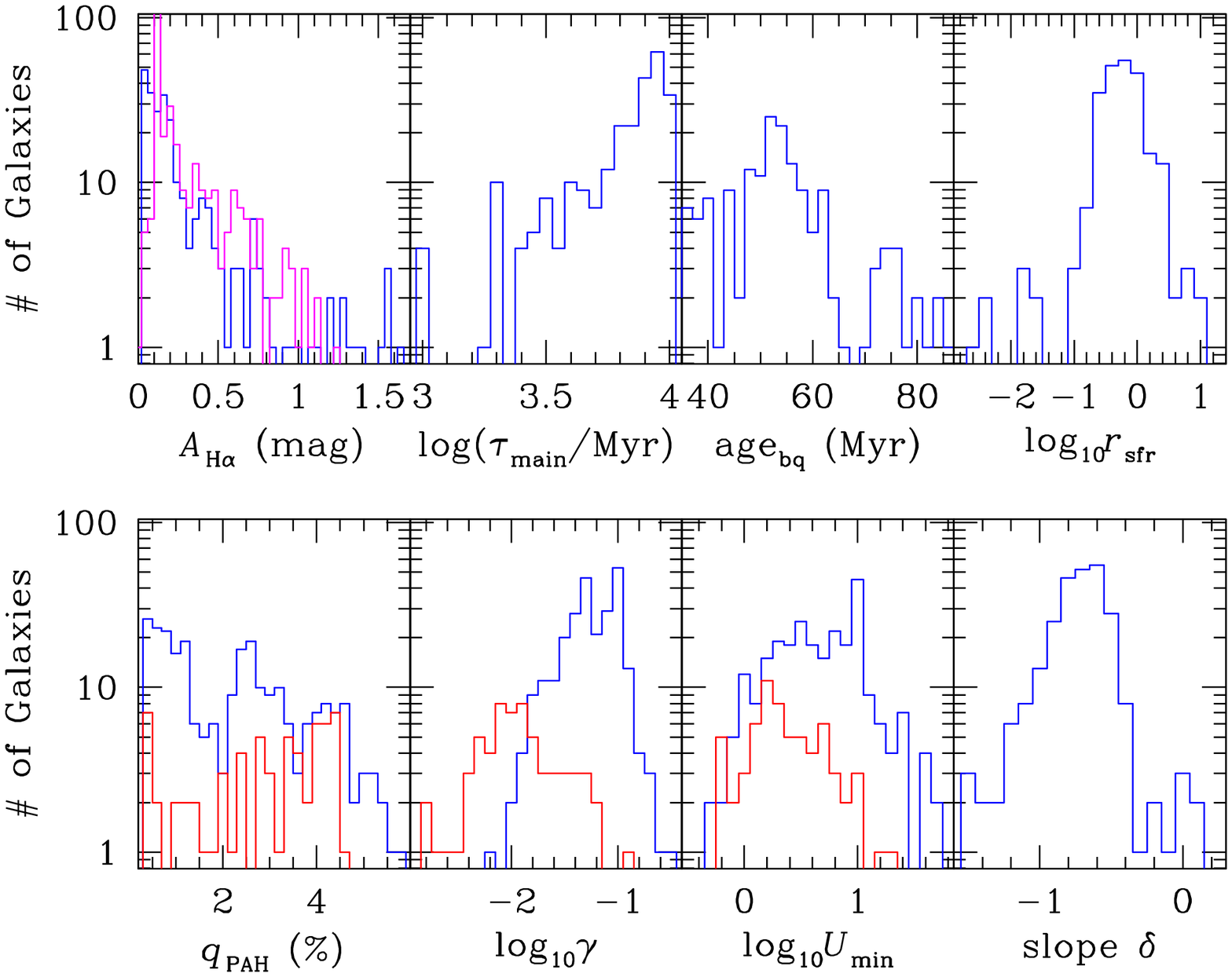}
 \caption{The distributions of Bayesian-based fit values for several key parameters are plotted in blue.  The red histograms overlaid for the dust parameters $\gamma$, $q_{\rm PAH}$, and $U_{\rm min}$ show the values from the spectral analysis of the SINGS sample of nearby galaxies in \cite{draine2007}.  The magenta histogram for $A_{{\rm H}\alpha}$ shows the distribution for the 11HUGS survey which has large overlap with LVL \citep{lee2009}.  
 Table~\ref{tab:medians} provides the median values and the $25^{\rm th}-75^{\rm th}$ percentile ranges for these parameters.}
 \label{fig:histograms}
\end{figure*}

Figure~\ref{fig:histograms} provides the distributions of Bayesian-based values for several main parameters in the CIGALE fits.  We include in Figure~\ref{fig:histograms} comparisons with results from other surveys.  The magenta histogram for $A_{{\rm H}\alpha}$ attenuations derives from the \cite{lee2009} 11HUGS survey of $\sim300$ star-forming galaxies, and the red histograms for $\gamma$, $q_{\rm PAH}$, and $U_{\rm min}$ come from the \cite{draine2007} analysis of 75 nearby star-forming galaxies from the SINGS survey \citep{kennicutt2003}.  The median values and the $25^{\rm th}-75^{\rm th}$ percentile ranges for these key parameters are provided in Table~\ref{tab:medians}.  The distribution of LVL $A_{{\rm H}\alpha}$ values displayed in Figure~\ref{fig:histograms} is qualitatively similar to the distribution of values for the 11HUGS sample, and the sample medians are quite close: 0.16~mag for LVL and 0.18~mag for 11HUGS.  An additional comparison point is that from the sample of $\sim$700,000 $z<0.3$ galaxies of \cite{salim2016} \citep[see also][]{salim2020}), which also has a median $A_{{\rm H}\alpha}$ value of 0.16~mag (where we have assumed $A_{{\rm H}\alpha} = 0.783 A_V$ from the \cite{draine2003} reddening curve).

\begin{deluxetable}{lrr}
\tabletypesize{\scriptsize}
\tablecaption{SED Fit Medians and $25^{\rm th}-75^{\rm th}$ Percentile Ranges\label{tab:medians}}
\tablewidth{0pc}
\tablehead{
\colhead{Parameter} &
\colhead{Median} &
\colhead{$25^{\rm th}-75^{\rm th}$} 
\\
\colhead{} &
\colhead{} &
\colhead{Percentile Range} 
}
\startdata
\hline
$\chi^2_{\rm reduced}$           &   0.81 & 0.41\\
$A_{{\rm H}\alpha}$ (mag)        &   0.16 & 0.14\\
Slope modifier $\delta$          &$-$0.71 & 0.13\\
$\log_{10}(\tau_{\rm main}$/Myr) &   3.90 & 0.11\\
$r_{\rm sfr}$                    &   0.52 & 0.33\\
$\log_{10}\gamma$                &$-$1.26 & 0.19\\
$q_{\rm PAH}$ (\%)               &   2.16 & 1.15\\
$\log_{10}U_{\rm min}$           &   0.68 & 0.31\\
age$_{\rm bq}$ (Myr)             &   47.  & 15.\\
\hline
\enddata
\end{deluxetable}

Two additional parameters that are allowed to vary are \texttt{alpha} and \texttt{E\_BV\_factor} (see Table~\ref{tab:parameters}).  The majority (56.2\%) of the best-fit solutions for the radiation field power-law exponent \texttt{alpha} is 2.0, matching the fixed value adopted by \cite{draine2007} who find their fits to nearby galaxy SEDs are largely insensitive to the particular choice for \texttt{alpha}.  Also consistent with previous efforts are the fit results for \texttt{E\_BV\_factor}: 81\% of our best-fit values for \texttt{E\_BV\_factor} are either 0.25 or 0.50, values which bracket the recommended value of 0.44 recommended by \cite{calzetti2000}.

The panels in Figure~\ref{fig:histograms} presenting the CIGALE-based fit distributions for the dust parameters $q_{\rm PAH}$, $\gamma$, and $U_{\rm min}$ are in reasonable agreement with published distributions for normal star-forming galaxies; the (non-CIGALE-based) fit values for these three parameters from the analysis of SINGS galaxies \citep{kennicutt2003} are similar to what we find for the LVL sample with CIGALE fits \citep{draine2007,dale2012}.  One clear discrepancy, however, is that the LVL distribution for $\gamma$ is shifted to larger values than what \cite{draine2007} find for the SINGS sample.  The dust parameter $\gamma$ is an indicator of the fraction of the interstellar dust that is heated by intense starlight such as in photodissociation regions that are associated with star-forming regions, i.e., not the bulk of the dust that is more gently heated throughout the diffuse interstellar medium of a galaxy.  This discrepancy is likely due to the LVL sample having a larger fraction of bursty star-forming dwarf galaxies, e.g., 58 LVL galaxies have SED fits with $r_{\rm SFR}$ values greater than unity (see Figure~\ref{fig:histograms}). 

We caution that any detailed comparisons between the results from our work and those from other efforts is limited by the different approaches to the SED fitting as well as the samples and data involved.  For example, the $A_{{\rm H}\alpha}$ attenuations from \cite{lee2009} are ultimately derived from a measured Balmer decrement.  In addition, \cite{draine2007} do not use an SED fitting procedure that balances the energy absorbed by dust at shorter wavelengths with the dust emission observed at longer wavelengths.  Also, CIGALE adopts $U_{\rm max}=10^7$ for the maximum value of the radiation field that heats the dust in units of the local interstellar radiation field \citep{dale2001} whereas only $\sim8$\% of the fit values drawn from \cite{draine2007} employ the same maximum $U$ (the remainder of the \cite{draine2007} fit values portrayed in Figure~\ref{fig:histograms} are based on $U_{\rm max}=10^6$).  Finally, \cite{janowiecki2017} study extreme starbursting blue compact dwarf galaxies and they adopt an exponentially-decaying star formation history for their CIGALE SED fits while we assume a (similar but) delayed exponential star formation history.

Another approach to assessing the robustness of the SED fits is to compare the fit parameters to those obtained when mock SEDs are fit, since the physical properties of the mock systems are precisely known.  CIGALE has built-in functionality for executing mock SED fits; for a given object the fluxes for the best fit model are modified by adding values drawn from Gaussian distributions with 1$\sigma$ values that match the flux uncertainties.  Based on the level of agreement between the mock fit parameter values and our best fit parameter values, the inference is that for our work there are three parameters that are less well constrained than the remaining parameters: $\gamma$, $\delta$, and $\alpha$.  For all three of these parameters, a comparison of the mock and best fit parameters values yields a Spearman's rank correlation coefficient less than 0.7 and an $r^2$ value for a linear regression less than 0.15.  This result is similar to those from other CIGALE-based galaxy SED fitting studies where the shapes of the dust SED ($\alpha$) and the assumed attenuation curve ($\delta$) are the least reliable fit parameters \citep[e.g.,][]{giovannoli2011,boquien2012}.

\begin{figure}
 \plotone{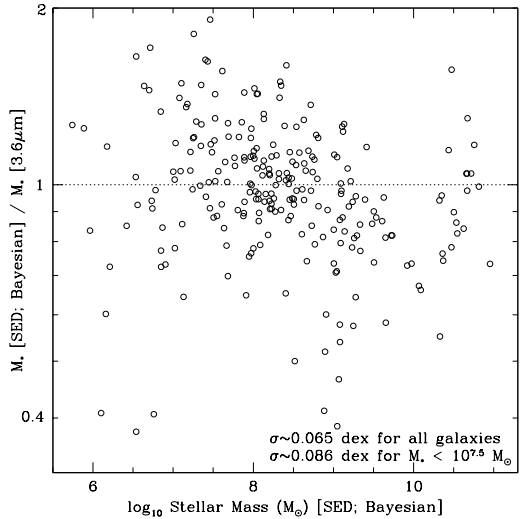}
 \caption{Comparison of CIGALE SED-based stellar masses with those derived from 3.6~\m\ infrared photometry.  The 3.6~\m\--based stellar masses are computed using a mass-to-light ratio that depends on the specific star formation rate following Appendix~A.4 of \cite{leroy2019}.  The horizontal dotted line indicates unity for the mass ratio.}
 \label{fig:stellar_mass}
\end{figure}

Figure~\ref{fig:stellar_mass} provides a comparison of the (Bayesian) SED-based stellar masses with those from single-band photometry assuming a mass-to-light ratio that depends on the specific star formation rate following Appendix~A.4 of \cite{leroy2019} \citep[see also][]{mcgaugh2014,meidt2014,hunt2019}.  There is general agreement between the stellar mass measures; the median ratio of the 3.6~\m-based-to-SED-based stellar masses agree within 0.01~dex of unity, and the dispersion (i.e., random scatter) is 0.065~dex.  There is a weak trend in Figure~\ref{fig:stellar_mass} as a function of the single-band-based stellar mass (Spearman's rank correlation coefficient of $-$0.29), and the scatter is somewhat enhanced for the lower mass systems; the scatter for galaxies with $\log_{10} M_*/M_\odot < 7.5$---galaxies for which the photometry is more challenging given their fainter flux levels---is 0.086~dex.

\begin{figure*}
 \includegraphics[height=13cm,trim={0 0 0 5cm},clip]{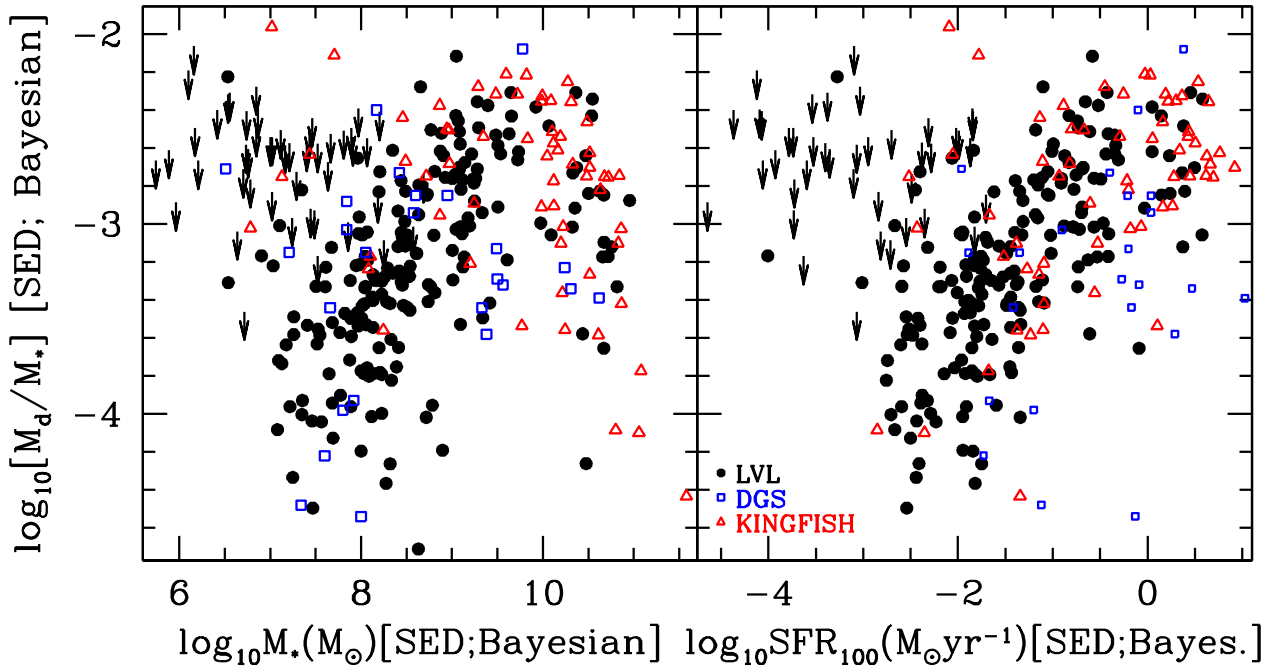}
 \caption{Left: Dust-to-stellar mass as a function of stellar mass.  Filled circles indicate our results for the LVL sample, the open red triangles are from the CIGALE SED fits for the KINGFISH sample from \cite{hunt2019}, and the open blue squares represent the SED-based values from the Dwarf Galaxy Sample presented in \cite{remyruyer2015}.  Right: Dust-to-stellar mass as a function of star formation rate averaged over the past 100~Myr, as inferred from the SED fitting (the star formation rates for the DGS sample are based on H$\alpha$$+$TIR).}
 \label{fig:dust_v_stellar}
\end{figure*}

The left panel of Figure~\ref{fig:dust_v_stellar} shows the trend in dust-to-stellar mass as a function of stellar mass, and the right panel shows the same ratio as a function of the star formation rate averaged over the past 100~Myr.  The median logarithmic ratio in $M_{\rm dust}/M_*$ for all LVL galaxies with secure dust masses is $-3.2$ with a dispersion of 0.36~dex.  The fit values from the \cite{hunt2019} CIGALE analysis of the KINGFISH sample of 61 nearby galaxies \citep{kennicutt2011} are included as open red triangles; clearly the KINGFISH sample is skewed, on average, to larger stellar masses and larger star formation rates compared to LVL.  We also provide data from the \cite{remyruyer2015} \citep[see also][]{hunt2019} analysis of the Dwarf Galaxy Sample (DGS); the DGS data are more scattered.  There are positive correlations in Figure~\ref{fig:dust_v_stellar} for the LVL sample, with Spearman's rank correlation coefficient values $\rho$ of 0.58 and 0.65, respectively, for the lefthand and righthand panels.  The trend seen in the righthand panel for the LVL sample echoes that seen for the more actively star-forming KINGFISH sample ($\log_{10}SFR_{100} (M_\odot~{\rm yr}^{-1})>-3$); Spearman's rank correlation coefficient $\rho=0.40$).

\begin{figure*}
 \plotone{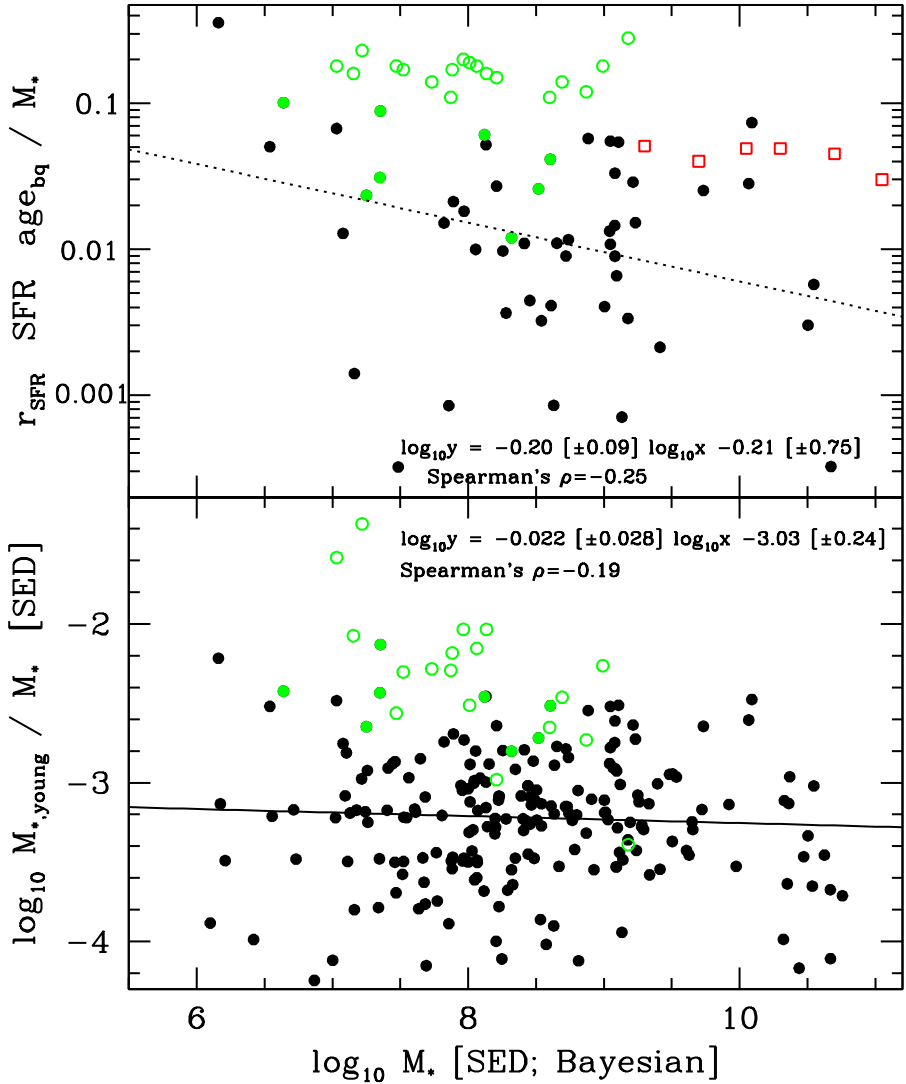}
 \caption{Top: The fraction of stellar mass in the modeled late burst of star formation compared to the total stellar mass in the galaxy, as a function of total stellar mass.  Bottom: the ratio of the mass in the young stellar population ($<10$~Myr) to the total mass of the stellar population ($>10$~Myr).  Filled circles correspond to LVL galaxies; those in green indicate blue compact dwarf galaxies in the LVL sample.  The green open circles are from the \cite{janowiecki2017} survey of blue compact dwarf galaxies and the red open squares are from the \cite{bergvall2016} study of SDSS starbursting galaxies.  Note that the upper panel is restricted to late bursts and not late quenching episodes (i.e., $r_{SFR}>1$).}
 \label{fig:young_mass_fractions}
\end{figure*}

The top panel of Figure~\ref{fig:young_mass_fractions} shows how the modeled fraction of stellar mass that derives from the late burst of star formation varies with total stellar mass, and the bottom panel displays the modeled ratio of stellar mass currently appearing in young stars ($<10~$Myr) to the total stellar mass, again as a function of the total stellar mass.  Both panels in Figure~\ref{fig:young_mass_fractions} show a modest decreasing trend with total stellar mass.  The burst fraction (top panel) is larger than the current young stellar mass fraction (bottom panel) due to young massive stars dying more quickly than older stars.  We note that a typical LVL galaxy has $\log_{10} M_{*,<10{\rm Myr}}/M_* \sim -3$, which roughly corresponds to the current star formation being the same as its average past value over the age of the disk.  To promote a fair comparison with the work of \cite{janowiecki2017}, we have restricted the upper panel to only LVL galaxies where the late episode is a burst ($r_{\rm SFR}>1$) and not a quenching event ($r_{\rm SFR}<1$).  Nonetheless, the comparison with \cite{janowiecki2017} still is not perfect, as their delayed star formation history adopts an exponentially-decaying late burst whereas our delayed star formation history uses a constant late star formation rate that lasts for \texttt{age\_bq} (see Equation~\ref{eqn:sfr}).  The values from the \cite{janowiecki2017} study lie at relatively higher values of burst fractions and higher values of young-to-total stellar mass ratios, not surprising given the nature of their sample of starbursting blue compact dwarf galaxies.  It is interesting to note that the eight LVL galaxies that have been classified as blue compact dwarfs \citep{gildepaz2003,kennicutt2008,lee2009b} also appear at elevated values in both panels of Figure~\ref{fig:young_mass_fractions}.  Another useful comparison sample provided in the top panel of Figure~\ref{fig:young_mass_fractions} is that of \cite{bergvall2016}.  They define starburst galaxies as having a birthrate parameter $b = SFR /\left< SFR \right> \geq 3$, and find that 1\% of star-forming galaxies SDSS DR7 galaxies qualify as starbursting.  Over their starburst subsample's stellar mass range of $\log_{10} M_*/M_\odot \sim 9-11.5$, they find higher burst fractions than we do for the LVL sample over a broader stellar mass range ($\log_{10} M_*/M_\odot \sim 6-11$): their median trend is within $\sim 0.02-0.05$ whereas we find a fitted trend with amplitude $\sim 0.004-0.04$.  Interestingly, the strongest decreasing trend with mass found in \cite{bergvall2016} is for post-starburst galaxies.

\begin{figure*}
 \includegraphics[height=13cm,trim={0 0 0 5cm},clip]{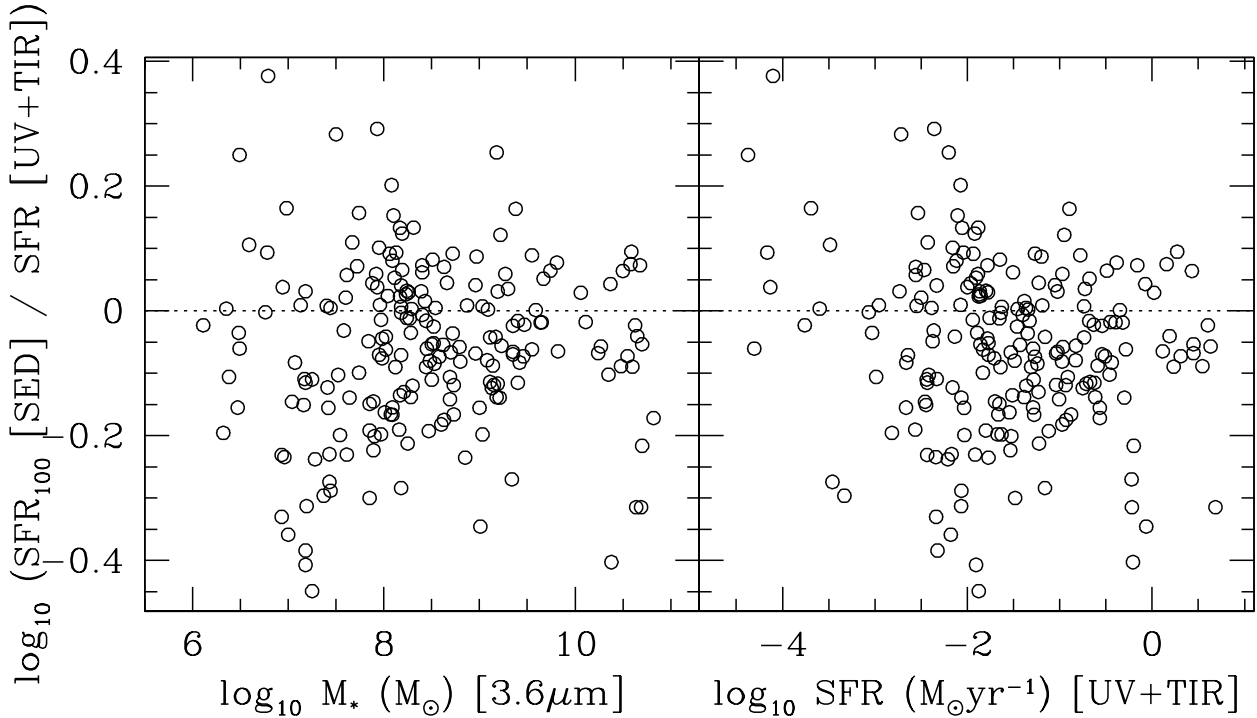}
 \caption{The (Bayesian-based) star formation rate from the SED fit averaged over 100~Myr compared to the star formation rate derived from hybrid indicators of ultraviolet + total infrared  \citep{hao2011}. The lefthand panel is plotted as a function of the 3.6~\m-based stellar mass and the righthand panel is shown as a function of hybrid-based star formation rates.  The horizontal dotted line indicates unity or perfect agreement between the star formation rates.}  
 \label{fig:sfr}
\end{figure*}

Figure~\ref{fig:sfr} displays how the star formation rate estimates compare between those derived on 100~Myr timescales from the SED fitting and those derived from the ``empirical'' ultraviolet + total infrared measurements (which also traces the star formation rate on 100~Myr timescales; \citealt{hao2011,kennicutt2012}).  There is no dependence on the ratio of modeled-to-empirical star formation rate indicators as a function of either stellar mass or star formation rate, and the median of the distribution is $-0.055$~dex with a scatter of 0.13~dex.  In other words, the modeled star formation rates for the bulk of the sample are about 88\% of the empirical values, on average.  The scatter is driven by outliers with ordinate ($y$-axis) values more than $\sim0.2$~dex from perfect agreement ($| \log 10 (SFR_{100}{\rm [SED]} / SFR{\rm [UV+TIR]} | > 0.2$).  Inspection of the CIGALE fits shows that a large number of these outliers have only upper limits for the MIPS infrared fluxes (and thus their $UV+TIR$ star formation rates are compromised).  For reference, \cite{salim2016} find good one-to-one agreement and a scatter of 0.2~dex when comparing their CIGALE-based star formation rates with those from SDSS spectroscopy.  As a reminder, our sample is dominated by dwarf galaxies, which we define here as $\log_{10} M_*/M_\odot < 9$; fully 75\% of LVL galaxies satisfy this dwarf status criterion.  In contrast, the SDSS-based work of \cite{salim2016} has only 1.5\% of their sample below this dwarf galaxy stellar mass threshold.  The LVL sample is also dominated by low-metallicity systems, which helps to explain the (small) discrepancy noted above between the SED-based and the $UV+TIR$-based values.  As noted by \cite{kennicutt2012}, lower metallicity systems have enhanced ultraviolet emissivity for a given star formation rate ($\sim 0.07$~dex for $Z=0.1Z_\odot$).

\begin{figure*}
 \includegraphics[height=13cm,trim={0 0 0 5cm},clip]{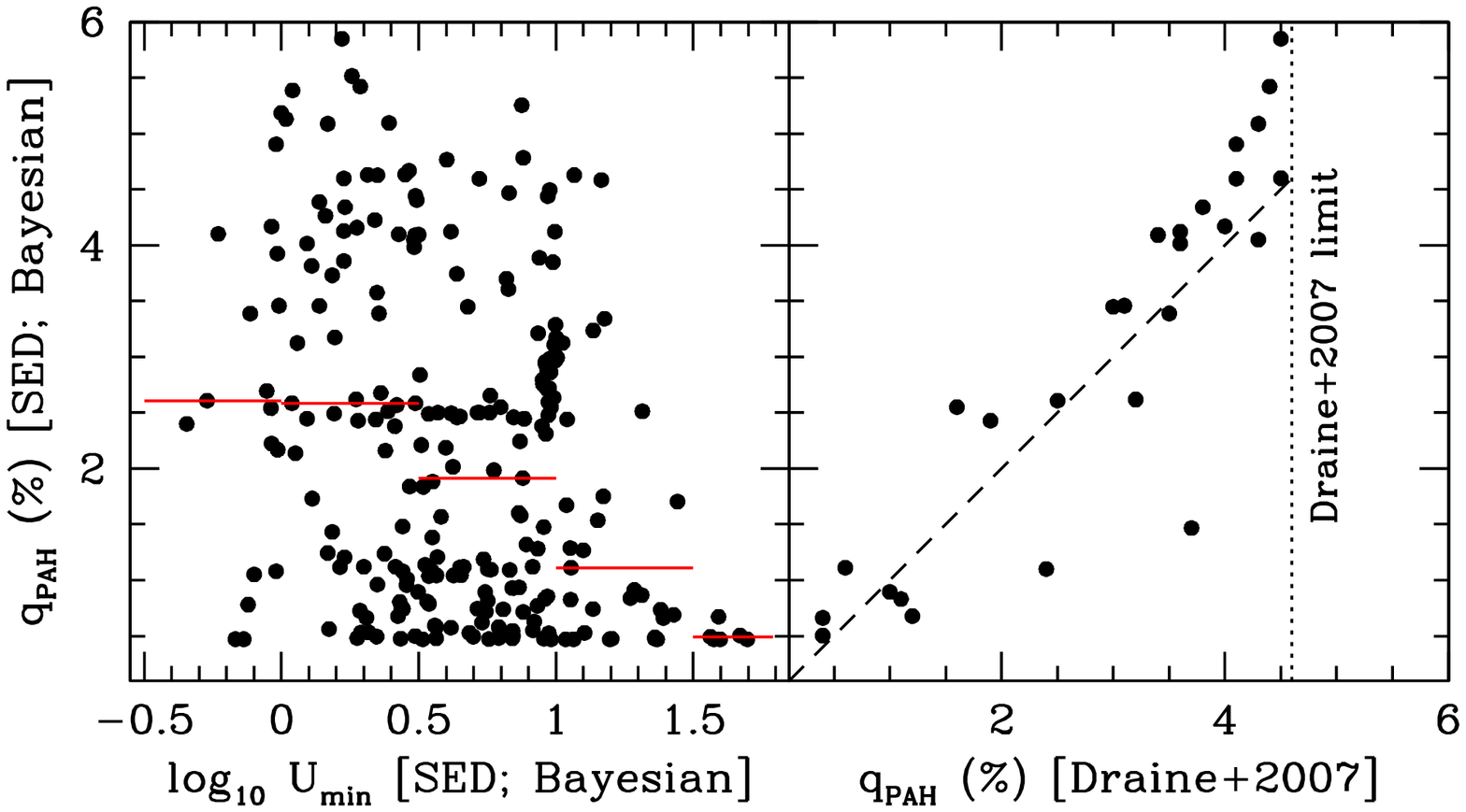}
 \caption{The fractional mass abundance of PAHS versus minimum interstellar radiation field intensity (left) and the fractional PAH mass abundance from the SED fitting work of \cite{draine2007} for the subset of the LVL sample that overlaps with the SINGS sample (right).  The median values for several bins are provided in the lefthand panel, the dotted line in the righthand panel indicates the maximum possible value for the \cite{draine2007} SED fits, and the dashed line in the righthand panel indicates the 1-to-1 relation.}
 \label{fig:qpah}
\end{figure*}

A final analysis of the SED fits is provided in Figure~\ref{fig:qpah}, which displays the fractional mass abundance of PAHs as a function of minimum interstellar radiation field that heats the dust in the diffuse ISM, as inferred from the SED fits, and the fractional PAH mass abundance for the subset of LVL galaxies that are in the SINGS sample and were fit by \cite{draine2007}.  Previous studies of PAH abundances have shown that the strength of PAH emission is sensitive to the hardness of the interstellar radiation field, the metal abundance, molecular gas surface density, star formation activity level (star formation rate, dust temperature, etc.), or some combination thereof \citep[e.g.,][]{engelbracht2005,madden2006,wu2006,draine2007,engelbracht2008,dale2009,sandstrom2010,wu2010,remyruyer2015,chastenet2019,galliano2021,dale2022}.  The LVL sample is a useful testbed for examining how dwarf galaxies and their generally lower metallicities and harder radiation fields affect PAHs.  The lefthand panel of Figure~\ref{fig:qpah} shows that $q_{\rm PAH}$ decreases as a function of the intensity of radiation field that heats the diffuse interstellar medium, suggesting that the radiation field plays a role in processing (e.g., ionizing, destroying) small dust grains/molecules.  The righthand panel of Figure~\ref{fig:qpah} indicates that our SED fits provide fractional PAH mass abundances that are consistent with other independent SED-fitting efforts.

\section{Summary} \label{sec:summary}
We have carried out a detailed study of the spectral energy distributions for the 258 galaxies in the Local Volume Legacy program.  This volume-limited sample presents a statistically representative view of the overall galaxy population in the local universe; approximately three-quarters of the sample is classified as dwarf galaxies with $\log_{10} M_*/M_\odot < 9$, a portion of parameter space that is rarely studied with statistically large samples of star-forming galaxies.  The data set utilized here involves fluxes at 26 different wavelengths spanning the far-ultraviolet through the far-infrared.  The CIGALE fitting code was used to characterize each galaxy's physical properties assuming a delayed star formation history along with a late starburst/quenching episode.  Reasonable fits ($\chi^2_{\rm reduced}<3$) are obtained for 94\% of the sample, and the SED-derived stellar masses agree with those traditionally derived from near-infrared photometry to within better than 0.1~dex.  We find that the stellar mass fraction arising from the modeled late bursts decreases for increasingly massive galaxies, consistent with the notion that low-mass dwarf galaxies are more likely to experience bursty star formation histories than massive star-forming galaxies \citep[e.g.,][]{weisz2012}.  We also find that the SED-derived star formation rates for LVL galaxies are on average 88\% the values obtained from ``empirical'' hybrid star formation rate indicators.  Finally, the SED fits indicate that the PAH fractional dust mass decreases for increasing values of the radiation field that heats the dust dispersed throughout the diffuse interstellar medium.



\acknowledgments
{
This work is based in part on observations made with the \textit{Spitzer Space Telescope}, which is operated by the Jet Propulsion Laboratory, California Institute of Technology under a contract with NASA.

This publication makes use of data products from the \textit{Wide-field Infrared Survey Explorer (WISE)}, which is a joint project of the University of California, Los Angeles, and the Jet Propulsion Laboratory/California Institute of Technology, funded by the National Aeronautics and Space Administration.

This work is based in part on observations made with the \textit{Galaxy Evolution Explorer (GALEX)}. \textit{GALEX} is a NASA Small Explorer, whose mission was developed in cooperation with the Centre National d'Etudes Spatiales (CNES) of France and the Korean Ministry of Science and Technology. \textit{GALEX} is operated for NASA by the California Institute of Technology under NASA contract NAS5-98034.

Funding for the SDSS and SDSS-II has been provided by the Alfred P. Sloan Foundation, the Participating Institutions, the National Science Foundation, the U.S. Department of Energy, the National Aeronautics and Space Administration, the Japanese Monbukagakusho, the Max Planck Society, and the Higher Education Funding Council for England.

The SDSS is managed by the Astrophysical Research Consortium for the Participating Institutions. The Participating Institutions are the American Museum of Natural History, Astrophysical Institute Potsdam, University of Basel, University of Cambridge, Case Western Reserve University, University of Chicago, Drexel University, Fermilab, the Institute for Advanced Study, the Japan Participation Group, Johns Hopkins University, the Joint Institute for Nuclear Astrophysics, the Kavli Institute for Particle Astrophysics and Cosmology, the Korean Scientist Group, the Chinese Academy of Sciences (LAMOST), Los Alamos National Laboratory, the Max-Planck-Institute for Astronomy (MPIA), the Max-Planck-Institute for Astrophysics (MPA), New Mexico State University, Ohio State University, University of Pittsburgh, University of Portsmouth, Princeton University, the United States Naval Observatory, and the University of Washington.

This publication makes use of data products from the Two Micron All Sky Survey, which is a joint project of the University of Massachusetts and the Infrared Processing and Analysis Center/California Institute of Technology, funded by the National Aeronautics and Space Administration and the National Science Foundation.

MB gratefully acknowledges support by the ANID BASAL project FB210003 and from the FONDECYT regular grant 1211000.
}




\appendix
\clearpage

Table~\ref{tab:fluxes} provides the photometry corrected for neither Galactic nor intrinsic extinction.  5$\sigma$ upper limits are included where available.  
The fit values are found in Table~\ref{tab:fits}.  Best-fit values are presented for $\chi^2_{\rm reduced}$ and Bayesian fit values and their corresponding uncertainties are presented for the remaining parameters.  $SFR_{100}$ is the star formation rate averaged over the past 100~Myr.  Masses assume the distances presented in \cite{dale2009}.

\startlongtable
\begin{deluxetable}{lrrrrrrrrr}
\tablenum{4}
\def\a{\tablenotemark{a}}
\def\p{$\pm$}
\floattable
\tabletypesize{\tiny}
\tablecaption{Spatially-Integrated Photometry Corrected for Neither Galactic Nor Intrinsic Extinction}
\tablewidth{0pc}
\tablehead{
\colhead{} &
\colhead{\GALEX} &
\colhead{\GALEX} &
\colhead{}      &
\colhead{}      &
\colhead{}      &
\colhead{}      &
\colhead{\SDSS} &
\colhead{\SDSS} &
\colhead{\SDSS} 
\\
\colhead{Filter} &
\colhead{FUV} &
\colhead{NUV} &
\colhead{{\it U}}  &
\colhead{{\it B}}  &
\colhead{{\it V}}  &
\colhead{{\it R$_{\rm C}$}} &
\colhead{{\it u}}  &
\colhead{{\it g}}  &
\colhead{{\it r}} 
\\
\colhead{$\bar{\lambda}$($\mu$m)} &
\colhead{0.154} &
\colhead{0.231} &
\colhead{0.361} &
\colhead{0.443} &
\colhead{0.551} &
\colhead{0.659} &
\colhead{0.356} &
\colhead{0.482} &
\colhead{0.626} 
\\
\colhead{$\Delta \lambda$($\mu$m)} &
\colhead{0.025} &
\colhead{0.073} &
\colhead{0.064} &
\colhead{0.099} &
\colhead{0.089} &
\colhead{0.159} &
\colhead{0.056} &
\colhead{0.127} &
\colhead{0.133} 
\\
\colhead{$A_\lambda/A_V$} &
\colhead{2.586} &
\colhead{2.994} &
\colhead{1.556} &
\colhead{1.291} &
\colhead{1.0}   &
\colhead{0.795} &
\colhead{1.589} &
\colhead{1.206} &
\colhead{0.854} 
}
\startdata
WLM          &280\p044E$-$2&313\p047E$-$2&\nodata      &100\p002E$-$1&145\p004E$-$1&173\p004E$-$1&\nodata      &\nodata      &\nodata      \\
NGC0024      &772\p121E$-$3&981\p149E$-$3&\nodata      &496\p013E$-$2&879\p024E$-$2&972\p026E$-$2&\nodata      &\nodata      &\nodata      \\
NGC0045      &291\p045E$-$2&356\p055E$-$2&642\p017E$-$2&128\p003E$-$1&199\p005E$-$1&226\p006E$-$1&\nodata      &\nodata      &\nodata      \\
NGC0055      &277\p043E$-$1&356\p055E$-$1&\nodata      &\nodata      &\nodata      &\nodata      &\nodata      &\nodata      &\nodata      \\
NGC0059      &\nodata      &\nodata      &928\p025E$-$3&212\p005E$-$2&380\p010E$-$2&492\p013E$-$2&\nodata      &\nodata      &\nodata      \\
ESO410-G005  &157\p026E$-$4&432\p066E$-$4&\nodata      &328\p009E$-$3&\nodata      &666\p018E$-$3&\nodata      &\nodata      &\nodata      \\
Sculptor-dE1 &197\p060E$-$6&854\p209E$-$5&\nodata      &570\p020E$-$4&\nodata      &113\p003E$-$3&\nodata      &\nodata      &\nodata      \\
ESO294-G010  &199\p047E$-$4&387\p059E$-$4&\nodata      &267\p007E$-$3&\nodata      &570\p015E$-$3&\nodata      &\nodata      &\nodata      \\
\enddata     
\label{tab:fluxes}
\end{deluxetable}

\clearpage
\startlongtable
\begin{deluxetable}{lrrrrrrrrr}
\tablenum{4}
\def\a{\tablenotemark{a}}
\def\p{$\pm$}
\floattable
\tabletypesize{\tiny} 
\tablecaption{Spatially-Integrated Photometry Corrected for Neither Galactic Nor Intrinsic Extinction}
\tablewidth{0pc}
\tablehead{
\colhead{} &
\colhead{\SDSS}    &
\colhead{\SDSS}    &
\colhead{\it 2MASS}   &
\colhead{\it 2MASS}   &
\colhead{\it 2MASS}   &
\colhead{\Spitzer} &
\colhead{\Spitzer} &
\colhead{\Spitzer} &
\colhead{\Spitzer} 
\\
\colhead{Filter} &
\colhead{{\it i}}   &
\colhead{{\it z}}   &
\colhead{{\it J}}   &
\colhead{{\it H}}   &
\colhead{{\it K$_{\rm s}$}}  &
\colhead{IRAC} &
\colhead{IRAC} &
\colhead{IRAC} &
\colhead{IRAC} 
\\
\colhead{$\bar{\lambda}$($\mu$m)} &
\colhead{0.767} &
\colhead{0.910} &
\colhead{1.24}  &
\colhead{1.65}  &
\colhead{2.17}  &
\colhead{3.56}  &
\colhead{4.51}  &
\colhead{5.76}  &
\colhead{7.96}  
\\
\colhead{$\Delta \lambda$($\mu$m)} &
\colhead{0.135} &
\colhead{0.131} &
\colhead{0.163} &
\colhead{0.252} &
\colhead{0.264} &
\colhead{0.687} &
\colhead{0.872} &
\colhead{1.25}  &
\colhead{2.55}  
\\
\colhead{$A_\lambda/A_V$} &
\colhead{0.647}  &
\colhead{0.494}  &
\colhead{0.291}  &
\colhead{0.185}  &
\colhead{0.118}  &
\colhead{0.0466} &
\colhead{0.0297} &
\colhead{0.0201} &
\colhead{0.0246} 
}
\startdata
WLM          &\nodata      &\nodata      &237\p014E$-$1&306\p019E$-$1&115\p015E$-$1&876\p118E$-$2&626\p086E$-$2&410\p052E$-$2&508\p063E$-$2\\
NGC0024      &\nodata      &\nodata      &228\p012E$-$1&246\p013E$-$1&188\p012E$-$1&101\p013E$-$1&682\p100E$-$2&821\p115E$-$2&117\p015E$-$1\\
NGC0045      &\nodata      &\nodata      &204\p013E$-$1&226\p016E$-$1&165\p017E$-$1&199\p027E$-$1&126\p017E$-$1&655\p083E$-$2&168\p020E$-$1\\
NGC0055      &\nodata      &\nodata      &327\p016E$+$0&305\p015E$+$0&262\p013E$+$0&201\p027E$+$0&139\p019E$+$0&151\p018E$+$0&240\p029E$+$0\\
NGC0059      &\nodata      &\nodata      &833\p054E$-$2&112\p007E$-$1&630\p071E$-$2&328\p044E$-$2&222\p030E$-$2&156\p020E$-$2&117\p014E$-$2\\
ESO410-G005  &\nodata      &\nodata      &$<$808E$-$3  &$<$121E$-$2  &$<$147E$-$2  &108\p014E$-$2&451\p061E$-$3&612\p081E$-$3&451\p057E$-$3\\
Sculptor-dE1 &\nodata      &\nodata      &$<$497E$-$3  &$<$743E$-$3  &$<$909E$-$3  &$<$339E$-$4  &$<$509E$-$4  &$<$187E$-$3  &$<$205E$-$3  \\
ESO294-G010  &\nodata      &\nodata      &172\p022E$-$2&210\p033E$-$2&134\p039E$-$2&478\p064E$-$3&311\p042E$-$3&$<$188E$-$3  &$<$205E$-$3  \\
\enddata     
\end{deluxetable}

\clearpage
\startlongtable
\begin{deluxetable}{lrrrrrrrr}
\tablenum{4}
\def\a{\tablenotemark{a}}
\def\p{$\pm$}
\floattable
\tabletypesize{\tiny}
\tablecaption{Spatially-Integrated Photometry Corrected for Neither Galactic Nor Intrinsic Extinction}
\tablewidth{0pc}
\tablehead{
\colhead{} &
\colhead{\Spitzer} &
\colhead{\Spitzer} &
\colhead{\Spitzer} &
\colhead{\IRAS}    &
\colhead{\IRAS}    &
\colhead{\IRAS}    &
\colhead{\IRAS}    &
\colhead{}
\\
\colhead{Filter} &
\colhead{MIPS} &
\colhead{MIPS} &
\colhead{MIPS} &
\colhead{}     &
\colhead{}     &
\colhead{}     &
\colhead{}     &
\colhead{H$\alpha$}
\\
\colhead{$\bar{\lambda}$($\mu$m)} &
\colhead{23.8}  &
\colhead{72.5}  &
\colhead{157}   &
\colhead{11.6}  &
\colhead{23.9}  &
\colhead{61.5}  &
\colhead{101.9} &
\colhead{}  
\\
\colhead{$\Delta \lambda$($\mu$m)} &
\colhead{5.32} &
\colhead{21.3} &
\colhead{35.8} &
\colhead{6.07} &
\colhead{10.0} &
\colhead{30.5} &
\colhead{33.3} &
\colhead{} 
\\
\colhead{$A_\lambda/A_V$} &
\colhead{0.0226}   &
\colhead{0.00245}  &
\colhead{0.000472} &
\colhead{0.0375}   &
\colhead{0.0204}   &
\colhead{0.00356}  &
\colhead{0.00118}  &
\colhead{0.783} 
}
\startdata
WLM          &755\p033E$-$2&200\p014E$+$0&388\p047E$+$0&\nodata      &\nodata      &889\p177E$-$1&164\p032E$+$0&201\p058E$+$0\\
NGC0024      &120\p004E$-$1&222\p015E$+$0&712\p085E$+$0&\nodata      &163\p032E$-$1&158\p031E$+$0&466\p093E$+$0&117\p012E$+$0\\
NGC0045      &187\p007E$-$1&386\p027E$+$0&123\p014E$+$1&\nodata      &\nodata      &195\p039E$+$0&552\p110E$+$0&492\p057E$+$0\\
NGC0055      &628\p025E$+$0&125\p008E$+$2&253\p030E$+$2&133\p026E$+$0&625\p125E$+$0&769\p154E$+$1&174\p034E$+$2&609\p062E$+$1\\
NGC0059      &439\p018E$-$2&516\p040E$-$1&402\p060E$-$1&\nodata      &\nodata      &459\p091E$-$1&604\p120E$-$1&404\p054E$-$1\\
ESO410-G005  &$<$442E$-$3  &$<$598E$-$2  &$<$225E$-$1  &\nodata      &\nodata      &\nodata      &\nodata      &\nodata      \\
Sculptor-dE1 &$<$341E$-$3  &$<$463E$-$2  &$<$174E$-$1  &\nodata      &\nodata      &\nodata      &\nodata      &\nodata      \\
ESO294-G010  &$<$343E$-$3  &$<$463E$-$2  &$<$174E$-$1  &\nodata      &\nodata      &\nodata      &\nodata      &\nodata      \\
\enddata     
\tablecomments{\footnotesize The compact table entry format TUV$\pm$WXYEZ implies (T.UV$\pm$W.XY)$\times10^{\rm Z}$ in Jy for the broadband filters and $10^{-15}$~W~m$^{-2}$ for H$\alpha$.  Corrections for neither Galactic nor intrinsic extinction have been applied (see \S~\ref{sec:data}).  5$\sigma$ upper limits are provided for non-detections.  No color corrections have been applied.  The filter central wavelengths and widths are computed via Equations~1 and 2 of \cite{dale2017}.}
\tablenotetext{a}{\footnotesize The \Spitzer\ MIPS data for NGC~3034 suffer from saturation and thus the tabulated values are lower limits.}
\end{deluxetable}

\startlongtable
\begin{deluxetable*}{lccccccccccc}
\tablenum{5}
\tabletypesize{\scriptsize}
\tablecaption{SED Fit Values\label{tab:fits}}
\tablewidth{0pc}
\tablehead{
\colhead{Name} &
\colhead{$\chi^2_{\rm red}$} &
\colhead{$A_{{\rm H}\alpha}$} &
\colhead{$\delta$} &
\colhead{$\log_{10}\tau_{\rm main}$} &
\colhead{$r_{\rm SFR}$} &
\colhead{$\log_{10}\gamma$} &
\colhead{$q_{\rm PAH}$} &
\colhead{$\log_{10}U_{\rm min}$} &
\colhead{$\log_{10}SFR_{100}$} &
\colhead{$\log_{10}M_{\rm dust}$} &
\colhead{$\log_{10}M_*$}
\\
\colhead{} &
\colhead{} &
\colhead{(mag)} &
\colhead{} &
\colhead{(Myr)} &
\colhead{} &
\colhead{} &
\colhead{(\%)} &
\colhead{} &
\colhead{($M_\odot {\rm yr}^{-1}$)} &
\colhead{($M_\odot$)} &
\colhead{($M_\odot$)} 
}
\startdata
\hline
\hline
WLM          & 2.37&0.07$\pm$0.03&-0.6$\pm$ 0.5&3.97$\pm$0.05&0.35$\pm$0.10&-0.99$\pm$0.56&1.57$\pm$0.70& 0.58$\pm$0.29&-2.55$\pm$0.03& 3.77$\pm$0.28& 7.26$\pm$0.02\\
NGC0024      & 0.37&0.40$\pm$0.15&-0.5$\pm$ 0.4&3.73$\pm$0.15&0.44$\pm$0.16&-0.96$\pm$0.49&3.46$\pm$0.72& 0.14$\pm$0.27&-0.91$\pm$0.10& 6.46$\pm$0.25& 9.24$\pm$0.05\\
NGC0045      & 2.28&0.18$\pm$0.06&-0.9$\pm$ 0.4&3.97$\pm$0.05&0.78$\pm$0.13&-0.89$\pm$0.46&2.58$\pm$0.35& 0.04$\pm$0.30&-0.74$\pm$0.03& 6.61$\pm$0.25& 9.07$\pm$0.02\\
NGC0055      & 1.10&0.38$\pm$0.15&-0.6$\pm$ 0.5&3.88$\pm$0.14&2.29$\pm$1.05&-1.05$\pm$0.61&1.12$\pm$0.02& 0.42$\pm$0.21&-0.61$\pm$0.09& 6.57$\pm$0.21& 9.08$\pm$0.07\\
NGC0059      & 0.30&0.17$\pm$0.06&-0.7$\pm$ 0.5&3.32$\pm$0.06&1.22$\pm$0.36&-1.06$\pm$0.57&0.66$\pm$0.30& 1.39$\pm$0.21&-2.27$\pm$0.11& 3.92$\pm$0.24& 8.63$\pm$0.04\\
ESO410-G005  & 1.32&0.15$\pm$0.11&-0.8$\pm$ 0.5&3.93$\pm$0.09&0.00$\pm$0.03&-1.27$\pm$0.71&4.44$\pm$1.85& 0.97$\pm$0.71&-3.35$\pm$0.12&$<$ 4.15      & 6.76$\pm$0.05\\
Sculptor-dE1 & 1.39&0.29$\pm$0.51&-0.8$\pm$ 0.5&3.29$\pm$0.53&1.25$\pm$1.62&-1.26$\pm$0.71&2.99$\pm$2.17& 1.00$\pm$0.68&-4.46$\pm$1.23&$<$ 4.60      & 6.91$\pm$0.10\\
ESO294-G010  & 1.11&0.08$\pm$0.08&-0.8$\pm$ 0.5&3.61$\pm$0.18&0.14$\pm$0.27&-1.32$\pm$0.76&2.72$\pm$2.11& 0.97$\pm$0.70&-3.73$\pm$0.19&$<$ 4.02      & 6.78$\pm$0.08\\
\enddata
\end{deluxetable*}


 \clearpage
\bibliography{main.bib}




\end{document}